\documentclass[aps,prl,twocolumn,showpacs,amsmath,amssymb]{revtex4-1}

\usepackage{graphicx}
\usepackage{color}

\definecolor{Red}{rgb}{1.0,0.0,0.0}

\def\be#1{\begin{equation}\label{#1}}
\def\ee{\end{equation}}

\begin{document}

\title{Photon mass as a probe to extra dimensions}

\author{G. Alencar${}^a$}
\email{geova@fisica.ufc.br}
\author{C. R. Muniz${}^b$}
\email{celiomuniz@yahoo.com}
\author{R. R. Landim${}^a$}
\email{rrlandim@gmail.com}
\author{I. C. Jardim${}^a$}
\email{jardim@fisica.ufc.br}
\author{R. N. Costa Filho${}^a$}
\email{rai@fisica.ufc.br}

\affiliation{${}^a$Departamento de F\'{\i}sica, Universidade Federal
do Cear\'a, 60451-970 Fortaleza, Cear\'a, Brazil}
\affiliation{${}^b$Universidade Estadual do Cear\'a, Faculdade de Educa\c c\~ao, Ci\^encias e Letras do Sert\~ao Central - 
R. Epit\'acio Pessoa, 2554, 63.900-000  Quixad\'{a}, Cear\'{a},  Brazil.}

\date{\today}

\begin{abstract}

 In this manuscript we show that the geometrical localization mechanism implies a four dimensional mass for the photon. The consistence of the model provides a mass given exactly by $m_{\gamma}=\sqrt{R}/4$ where $R$ is the Ricci scalar. As a consequence, the cosmological photon has a mass related to the vacuum solution of the Einstein equation. At the present age of the universe we have a dS vacuum with $R=4\Lambda$, where Lambda is a positive cosmological constant. With this we find that $m_{\gamma}\approx 2\times 10^{-69}$ kg, which is below the present experimental upper bounds, and such correction may be observed in the next years with more precise measurements. By considering the value of $R$ inside some astrophysical sources and environments we find that the bound is also satisfied. The experimental verification of this mass, beyond pointing to the existence of extra dimensions, would imply in a fundamental change in cosmology, astrophysics and in particle physics since the same mechanism is valid for non-abelian gauge fields.

-

\end{abstract}

\maketitle
Compact extra dimensions has first been considered by Kaluza and Klein
in the $20$'s. In this model the only way to recover the four dimensional
physics is by considering a small compact extra dimension. The scenario
has changed in the $90$'s when Arkani et al proposed that the hierarchy
problem can be solved by a large extra dimensions \cite{ArkaniHamed:1998rs,Antoniadis:1998ig}. In this model the
metrics is factorable and the Einstein-Hilbert(EH) action becomes $\bar{M}_{pl}^{3}V\int d^{4}x\sqrt{g(x)}R(x)$. Here $\bar{M}_{pl}$ is the higher dimensional Planck mass, and $V$ is the volume of the compact space. Therefore, if a Tev scale for $\bar{M}_{pl}$ is considered we can get an effective four dimensional $M_{pl}=10^{18}$GeV if $V$ is large. Soon later Randall-Sundrum proposed a model, with a non-factorable metric, in which the extra dimension can be in
fact of infinity range \cite{Randall:1999ee,Randall:1999vf}. The idea is that a strong decreasing metric
can provide a finite integration even for an infinity range in the
extra dimension, and a consistent gravity theory is obtained over the
membrane. However, a drawback in the above models is that gauge fields
do not confine, failing to provide a consistent four dimensional observable
universe. More precisely, in a conformal coordinate the metrics is given
by $G_{MN}=e^{2A(z)}g_{MN}(x)$, where $A(z)$ is the warp factor,
$z$ and $x$ are the extra dimension and brane coordinates, respectively.
Before considering an specific form for the energy momentum tensor some
comments are important. The EH action is given by $S=2M^{3}\int d^{5}x\sqrt{-G}R(x,z)$
and for the above metric we know that the determinant of the metric is given by $\sqrt{-G}=e^{5A}\sqrt{-g(x)}$ and for the Ricci scalar
\begin{equation}\label{Ricci}
R(x,z)=e^{-2A}(R(x)-8A''-12A'^{2}).
\end{equation}
With this we see that the action contains a term given by
\begin{equation}
S\supset2\bar{M}^{3}\int dze^{3A}\int d^{4}x\sqrt{-g(x)}R(x),
\end{equation}
and the four dimensional EH action is recovered if $\int dze^{3A}$
is finite and this also relates the Planck mass in five and four dimensions
by $ M^{2}=\bar{M}^{3}\int dze^{3A}$.

In the RS setup we have a Minkowski vacuum with solution $A(z)=-\ln(k|z|+1)$.
This is obtained by considering a cosmological constant and two branes,
one with positive tension at the origin and one with negative tension
at the location $z_{c}$. Also, by imposing a Minkowski vacuum in four dimension, a fine tuning between the tensions
and the cosmological constant is needed. The above relation is then
given by
\begin{equation}
M^{3}=\frac{\bar{M}^{3}}{k}(1-\frac{1}{kz_{c}+1}).
\end{equation}
Here an interesting issue about this result is that it is valid for
$z_{c}\to\infty$, and the model becomes an alternative to compactification.
Next, Randall and Sundrum showed that the graviton zero mode is bound
to the brane.
Despite the fact that the model localize the graviton, it can not be applied to more complex gravitational scenarios since the RS brane is flat. For example,
the model is not suitable for a cosmological description, since the most recent observations indicate the existence of a positive cosmological constant.
Soon after the RS paper,  models with a non-vanish constant scalar curvature emerged \cite{Kim:1999ja, DeWolfe:1999cp}. 
In particular for the dS brane, which  describes the current expansion phase of the universe, the warp factor  is convergent.
Another advantage of these models is that no fine tuning between the brane tension and the bulk cosmological constant is needed.
These models localize the gravity, but the localization of other fields is not guaranteed, in particular the gauge fields.
For this case the action is given by
\begin{equation}
S_{A}=\frac{1}{4}\int d^{5}x\sqrt{-G}G^{MO}G^{NP}Y_{MN}Y_{OP}
\end{equation}
where $Y_{MN}=\partial_{M}X_{N}-\partial_{N}X_{M}$. Just like for
the gravity case, in order to obtain a well defined four dimensional
action the integration over the extra dimension must be finite. This
is attached by performing the separation of variables $X_{\mu}(x,z)=e^{-\frac{A}{2}}\psi(z)A_{\mu}(x)$
and the equations of motion(EOM) are given by
\begin{equation}
\partial_{\mu}\sqrt{g}F^{\mu\nu}=-m^{2}A^{\nu},
\end{equation}
with a Schr\"odinger like mass equation 
\begin{equation}
\psi''-V(z)\psi=m^{2}\psi.
\end{equation}
The prime means a $z$ derivative and $V(z)$ is given by 
\begin{equation}
V=\frac{A''}{2}+\frac{A'^{2}}{4}.
\end{equation}
With this, similarly to the gravity case, the five dimensional action
contains the term
\begin{equation}
S_{A}=\frac{1}{4}\int\psi^{2}dz\int d^{4}x\sqrt{-g}g^{\mu\nu}g^{\alpha\beta}F_{\mu\alpha}F_{\nu\beta},
\end{equation}
and the problem of obtaining a well defined action is resumed
to find a normalized solution with $\int\psi^{2}dz=1$. However,
differently from the gravity case the solution to the zero mode is not
normalizable. This is easily find since the above potential provides
the general analytical solution $\psi=e^{\frac{A}{2}}$ for the zero
mode. This is very similar to the gravity case but the integral do
not converge if asymptotically we have an AdS solution. In the search
for a solution to the gauge field localization which does not include
the addition of new degrees of freedom a new model has been proposed
in a series of papers \cite{Alencar:2014moa,Alencar:2014fga,Alencar:2015awa,Alencar:2015oka}. This model provides an analytical solution
given by $\psi=e^{A}$, which is a square integrable solution to the
mass equation valid for any warp factor recovering RS asymptotically.
The basic ingredient is the addition of a new term to the action given
by
\begin{equation}
S_{I}=-\frac{1}{32}\int d^{5}x\sqrt{-G}RG^{MN}X_{M}X_{N}.
\end{equation}

In the first version of the geometrical localization mechanism\cite{Alencar:2014moa}, the original RS model was considered when $R(x)=0$. Using Eq. (\ref{Ricci}) $R(x,z)=-e^{-2A}(8A''+12A'^{2})$, and a massless photon is obtained over the brane at least in first approximation.
Another interesting point about the interaction term is that it has
no free parameters and this will become crucial for determining the photon mass. From now on we will consider the full theory
and show that the gauge field is confined for arbitrary four dimensional
background. The proof is very similar to the flat case, but we give
it here for completeness. We must be careful since now we have $ds^{2}=e^{2A(z)}(g_{\mu\nu}(x)dx^{\mu}dx^{\nu}+dz^{2})$.
The equations of motion are 
\begin{equation}
\nabla_{M}(G^{MO}G^{NP}Y_{OP})=-\frac{1}{16}RG^{NO}X_{O},
\end{equation}
leading to the condition $\nabla_{N}(RG^{NO}X_{O})=0$, or
\begin{equation}\label{divergence}
e^{3A}\nabla_{\mu}(RX^{\mu})=-\nabla_{5}(e^{3A}RX_{5}).
\end{equation}

Since the gauge invariance is now broken by the interaction term we
have to show that a transversal gauge invariant zero mode is localized.
For this we must split the gauge field in longitudinal and transversal
components and show that they decouple. As mentioned before, in the previous version of the mechanism $R$
was independent of $x$ and the derivative on the left side of the
above identity did not act on it. However, we will see that this do
not spoil the decoupling. First we split our field as $X^{\mu}=X_{L}^{\mu}+X_{T}^{\mu}$,
where $L$ stands for longitudinal and $T$ stands for transversal
with $X_{T}^{\mu}=(\delta_{\nu}^{\mu}-\nabla^{\mu}\frac{1}{\Box}\nabla_{\nu})X^{\nu}\;\;\;;\;\;\;X_{L}^{\mu}=\nabla^{\mu}\frac{1}{\Box}\nabla_{\nu}X^{\nu}$.
Now we define $X_{5}=\Phi$ and the first thing we observe is that
(\ref{divergence}) will give us a relation between
the scalar field and the longitudinal part of $X^{\mu}$. We also
need the following identities 
\begin{eqnarray}
Y^{5\mu}&=&\partial X_{T}^{\mu}+\partial X_{L}^{\mu}-\partial^{\mu}\Phi\equiv\partial X_{T}^{\mu}+Y_{L}^{5\mu};\nonumber \\
Y_{L}^{\mu5}&=&\nabla^{\mu}\frac{1}{\Box}\nabla_{\nu}Y^{\nu5}. \label{identities}
\end{eqnarray}

Now considering the EOM for $N=5;\nu$ the set of equations are obtained:
\begin{equation}\label{M=5}
\nabla_{\mu}Y^{\mu5}+\frac{1}{16}e^{2A}R\Phi=0,
\end{equation}
and
\begin{eqnarray}
&&e^{A}\nabla_{\mu}Y^{\mu\nu}+\frac{1}{16}e^{3A}RX_{T}^{\nu}+\nonumber \\ 
&&+\nabla_{5}\left(e^{A}\partial X_{T}^{\nu}\right)+\nabla_{5}(e^{A}Y_{L}^{5\mu})+\frac{1}{16}e^{3A}RX_{L}^{\nu}=0. 
\end{eqnarray}
Using (\ref{divergence}),(\ref{identities}) and (\ref{M=5}) we get 
\begin{equation}
\nabla_{5}(e^{A}Y_{L}^{\mu5})=-\frac{1}{16}\nabla^{\mu}\frac{1}{\Box}\nabla_{5}(e^{3A}R\Phi)=-\frac{1}{16}e^{3A}RX_{L}^{\nu},
\end{equation}
and finally we can decouple the equation of motion for the transverse
part of the gauge field for arbitrary $R$
\begin{equation}
e^{A}\nabla_{\mu}Y^{\mu\nu}+\partial\left(e^{A}\partial X_{T}^{\nu}\right)+\frac{1}{16}e^{3A}RX_{T}^{\nu}=0.
\end{equation}

Finally performing the same transformation as before, or, $X_{\mu}(x,z)=e^{-\frac{A}{2}}\psi(z)A_{\mu}(x)$
we get the following equations of motion
\begin{equation}
\nabla_{\mu}(F^{\mu\nu})=-(m^{2}+R(x))A^{\nu},
\end{equation}
with a Schr\"odinger like mass equation 
\begin{equation}
\psi''-V(z)\psi=m^{2}\psi,
\end{equation}
but now with $V(z)$ given by 
\begin{equation}
V=A''+A'^{2}.
\end{equation}
The solution $\psi=ce^{A}$ for the zero mode and our four dimensional
action of the vector field is given by 
\begin{equation}
S_{A}=\frac{1}{4}\int c^{2}e^{2A}dz\int d^{4}x\sqrt{-g(x)}g^{\mu\nu}g^{\sigma\delta}F_{\mu\sigma}F_{\nu\delta}.
\end{equation}
 
Therefore the confining of the gauge field is reduced to the condition
$\int c^{2}e^{2A}dz=1$. It is important to point that for the warp
factor given before the above integral is convergent for any range
of the extra dimension and the four dimensional gauge action is recovered
for both RS models. Moreover, the above integral is also convergent
for any warp factor that recovers the RS metrics at the boundaries.
This is very powerful since the only condition is that we must consider
an AdS five dimensional vacuum to obtain a well defined theory. However, beyond the
above term we have the non-minimal coupling which will generate a
four dimensional contribution given by
\begin{equation}
S_{int}=-\frac{1}{32}\int c^{2}e^{2A}dz\int d^{4}x\sqrt{-g(x)}R(x)g^{\mu\nu}A_{\mu}A_{\nu}.
\end{equation}

Interestingly if we have a convergent solution to the gauge field this
term is necessarily confined and this spoils the four dimensional gauge
invariance. This results is a testable
prevision of the model. Note that, in the original RS model, the brane
vacuum is flat with $R(x)=0$ and the gauge invariance is recovered
for the vacuum. However, when we go to the next order we find a necessary
breaking of the gauge invariance throughout the above interaction.
Since we have no free parameters, the exact value of this mass can
be obtained, $m_{\gamma}=\sqrt{R}/4$. In the table \ref{table1} we consider the value of the
Ricci scalar inside some sources and we find that the mass obtained
for the photon is within the experimental bounds. The more
interesting and important case occurs when we consider that the four
dimensional universe is not flat but has a vacuum energy and $R=4\Lambda$,
where lambda is a positive cosmological constant. With this we get
that a cosmological photon propagating in the vacuum must have the
exact mass given by $m_{\gamma}\approx 2\times 10^{-69}$ kg. As far as we know this is the first model
in which the mass of the photon is a not a supposition but a necessary
ingredient. Moreover, since the mass is not a free parameter, this
provides a testable prevision of the model. We should point that the
above interaction, despite being very small, should have consequences
for astrophysics and stelar evolution. Since the above model is also
valid for non-abelian gauge fields\cite{Alencar:2015awa} we also must have consequences
to particle physics and we speculate that such kind interaction may
be observable in particle accelerator experiments.

To obtain values for the photon mass, $m_{\gamma}$, according to our geometric model of gauge field localization, we must consider some cosmological and astrophysical environments in order to calculate $R$ and then $m_{\gamma}=(\hbar/c)\sqrt{R}/4$, comparing such values with the respective constraints for the mass photon obtained from the current experimental and speculative inferences. To do this, we will suppose that the astrophysical medium is a perfect fluid with matter density $\rho$ and pressure $P$, with the vacuum being filled with an energy density which comes from the cosmological constant $\Lambda$. Thus the Ricci scalar is in the rest frame given by
\begin{equation}
R=\frac{8\pi G}{c^4}(\rho c^2-3P)+4\Lambda.
\end{equation}
For example, at the center of a neutron star, with $\rho\approx 5.0\times 10^{17}$ kg/m$^3$ and $P\approx 10^{32}$ Pa \cite{Haensel:2007yy}, we find $m_{\gamma}\approx 8\times10^{-48}$ kg. At the Sun core, $\rho\approx 1.5\times 10^{5}$ kg/m$^3$ and $P\approx 2\times10^{16}$ Pa \cite{Kenneth}, yielding $m_{\gamma}\approx 4\times 10^{-54}$ kg. Finally, in the vacuum, we have $R=4\Lambda$, and then $m_{\gamma}\approx 2\times 10^{-69}$ kg, for $\Lambda\approx 2\times 10^{-52}$/m$^2$ \cite{Carmeli:2001fb}.

It is interesting to establish a comparison, shown in the table \ref{table1}, of these and other values based on the model with the upper bounds ($m_{\gamma}\lesssim$) coming from both experimental procedures and theoretical estimates, according to \cite{Groom:2000in,Goldhaber:2008xy}.

\begin{table}[ht]
\caption{Photon mass values}
\centering
\begin{tabular}{|l|c|c|c|}\hline
{\bf Environments} & {\bf$m_{\gamma}\lesssim$} (kg) & {\bf$m^{0}_{\gamma}\thickapprox$} (kg) & {\bf$m^{\Lambda}_{\gamma}\thickapprox$}(kg) \\ \hline
Neutron star core & $-$ & $8\times10^{-48}$ & $8\times10^{-48}$\\ \hline
Terrestrial ionosphere & $4\times 10^{-49}$ & $2\times 10^{-57}$ & $2\times 10^{-57}$ \\ \hline
Jupiter magnetosphere & $7\times 10^{-52}$ & $10^{-65}$ & $10^{-65}$ \\ \hline
Sun core & $-$ & $4\times 10^{-54}$ & $4\times 10^{-54}$  \\ \hline
Sun magnetosphere & $2\times 10^{-54}$ & $2\times 10^{-72}$ & $2\times 10^{-69}$\\ \hline
Intergalactic medium & $10^{-62}$ & $2\times10^{-75}$ & $2\times10^{-69}$ \\ \hline
Cosmological vacuum & $-$ & $0$  & $2\times10^{-69}$ \\ \hline
\end{tabular}\label{table1}
\end{table}
Note that $m^{\Lambda}_{\gamma}$ and $m^0_{\gamma}$ are the photon masses calculated from the model with and without cosmological constant. The photon masses associated to the solar magnetosphere and intergalactic medium were calculated according to the energy density of the magnetic fields present in these scenarios, of $10^{-10}$ T and $10^{-13}$ T \cite{Arlen:2012iy}, respectively. It is worth also notice that all the obtained values are far below those upper limits.

\section*{Acknowledgment}
We acknowledge the financial support provided by Funda\c c\~ao Cearense de Apoio ao Desenvolvimento Cient\'\i fico e Tecnol\'ogico (FUNCAP), the Conselho Nacional de 
Desenvolvimento Cient\'\i fico e Tecnol\'ogico (CNPq) and FUNCAP/CNPq/PRONEX.

\providecommand{\href}[2]{#2}\begingroup\raggedright\endgroup
 
\end{document}